\documentclass[12pt]{article}

\usepackage{amsmath,amssymb,amsfonts,amsbsy}
\usepackage{graphicx}
\usepackage{wrapfig}
\begin{document}
\addcontentsline{toc}{subsection}{{Search for narrow pion-proton states in s-channel at EPECUR: experiment status}\\
{\it I.G. Alekseev}}

%%%%%%% please do not touch these! %%%%%%
\setcounter{section}{0}
\setcounter{subsection}{0}
\setcounter{equation}{0}
\setcounter{figure}{0}
\setcounter{footnote}{0}
\setcounter{table}{0}

\begin{center}
\textbf{SEARCH FOR NARROW PION-PROTON STATES IN S-CHANNEL AT EPECUR: EXPERIMENT STATUS}

\vspace{5mm}

\underline{I.G.~Alekseev}$^{\,1\,\dag}$,
V.A.~Andreev$^{\,2}$,
I.G.~Bordyuzhin$^{\,1}$,
P.Ye.~Budkovsky$^{\,1}$, 
D.A.~Fedin$^{\,1}$, 
Ye.A.~Filimonov$^{\,2}$, 
V.V.~Golubev$^{\,2}$, 
V.P.~Kanavets$^{\,1}$, 
L.I.~Koroleva$^{\,1}$, 
A.I.~Kovalev$^{\,2}$, 
N.G.~Kozlenko$^{\,2}$, 
V.S.~Kozlov$^{\,2}$, 
A.G.~Krivshich$^{\,2}$, 
B.V.~Morozov$^{\,1}$, 
V.M.~Nesterov$^{\,1}$, 
D.V.~Novinsky$^{\,2}$,
V.V.~Ryltsov$^{\,1}$, 
M.~Sadler$^{\,3}$,
A.D.~Sulimov$^{\,1}$,
V.V.~Sumachev$^{\,2}$, 
D.N.~Svirida$^{\,1}$, 
V.I.~Tarakanov$^{\,2}$ and 
V.Yu.~Trautman$^{\,2}$

\vspace{5mm}

\begin{small}
  (1) \emph{ITEP, Moscow} \\
  (2) \emph{PNPI, Gatchina} \\
  (3) \emph{ACU, Abilene, USA} \\
  $\dag$ \emph{E-mail: Igor.Alekseev@itep.ru}
\end{small}
\end{center}

\vspace{0.0mm} % Don't laugh: it does change the spacing!

\begin{abstract}
An experiment EPECUR, aimed at the search of the cryptoexotic non-strange member of the pentaquark antidecuplet, 
started its operation at a pion beam line of the ITEP 10~GeV proton synchrotron. 
The invariant mass range of the interest (1610-1770)~MeV will be scanned for a narrow state 
in the pion-proton and kaon-lambda systems in the “formation-type experiment. The scan in the 
s-channel is supposed to be done by the variation of the incident $\pi^-$-momentum and its measurement 
with the accuracy of up to 0.1\% with a set of 1~mm pitch proportional chambers located in the first 
focus of the beam line. The reactions under the study will be identified by a magnetless spectrometer 
based on wire drift chambers with a hexagonal structure. Because the background suppression in this 
experiment depends on the angular resolution, the amount of matter in the chambers and setup is 
minimized to reduce multiple scattering. The differential cross section of the elastic 
$\pi^-p$-scattering on a liquid hydrogen target in the region of the diffractive minimum 
will be measured with statistical accuracy 0.5\% in 1~MeV steps in terms of the invariant mass. 
For $K^0_S\Lambda^0$-production the total cross section will be measured with 1\% 
statistical accuracy in the same steps. An important byproduct of this experiment will be a very 
accurate study of $\Lambda$ polarization.
The setup was assembled and tested in December 2008 and in April 2009 
we had the very first physics run. About $0.5\cdot 10^9$ triggers were written to disk 
covering pion beam momentum range 940-1135 MeV/c.
The talk covers the experimental setup and the current status.
\end{abstract}

\vspace{7.2mm} 

   An interest to this experiment originated with the discovery in 2003 by the two experiments LEPS \cite{LEPS} and
DIANA \cite{DIANA} a new baryonic state $\theta^+$ with positive strangeness and very small width. Later 
appeared several strong results where the state was not seen \cite{CLAS} but recent results from LEPS \cite{LEPS1}
and DIANA \cite{DIANA1} still insist on the evidence for this resonance. Quantum numbers of $\theta^+$ are not
measured but it is believed that it belongs to pentaquark antidecuplet predicted in 1997 by D.~Diakonov, V.~Petrov
and M.~Polyakov \cite{DPP}. In this case there should also exist a non-strange neutral resonance P11 with mass near 
1700~MeV. Certain hints in favour of its presence were found in the modified PWA of GWU group \cite{GWU}
at masses 1680 and 1730~MeV 
\cite{PWAM}. Recently an indication for this narrow state was found in $\eta$-photoproduction on deuteron in GRAAL
\cite{GRAAL} and some other experiments. The structure observed has mass 1685~MeV and width $< 30$~MeV, which was
determined by the detector resolution.
   Our idea is to search for P11(1700) in formation-type experiment on a pion beam \cite{EPECUR}. 
Precise measurement of the beam momentum and fair statistics will allow us to do a scan with unprecedented
invariant mass resolution. We plan to measure differential cross sections of the reactions $\pi^-p\to\pi^-p$ and 
$\pi^-p\to K^0_S\Lambda^0$ with high statistics and better than a MeV invariant mass resolution. If the resonance 
does exist our experiment will provide statistically significant result and we will measure its width with the
precision better than 0.7~MeV.

\begin{figure}[t!]
    \centering
    \includegraphics[width=0.9\textwidth]{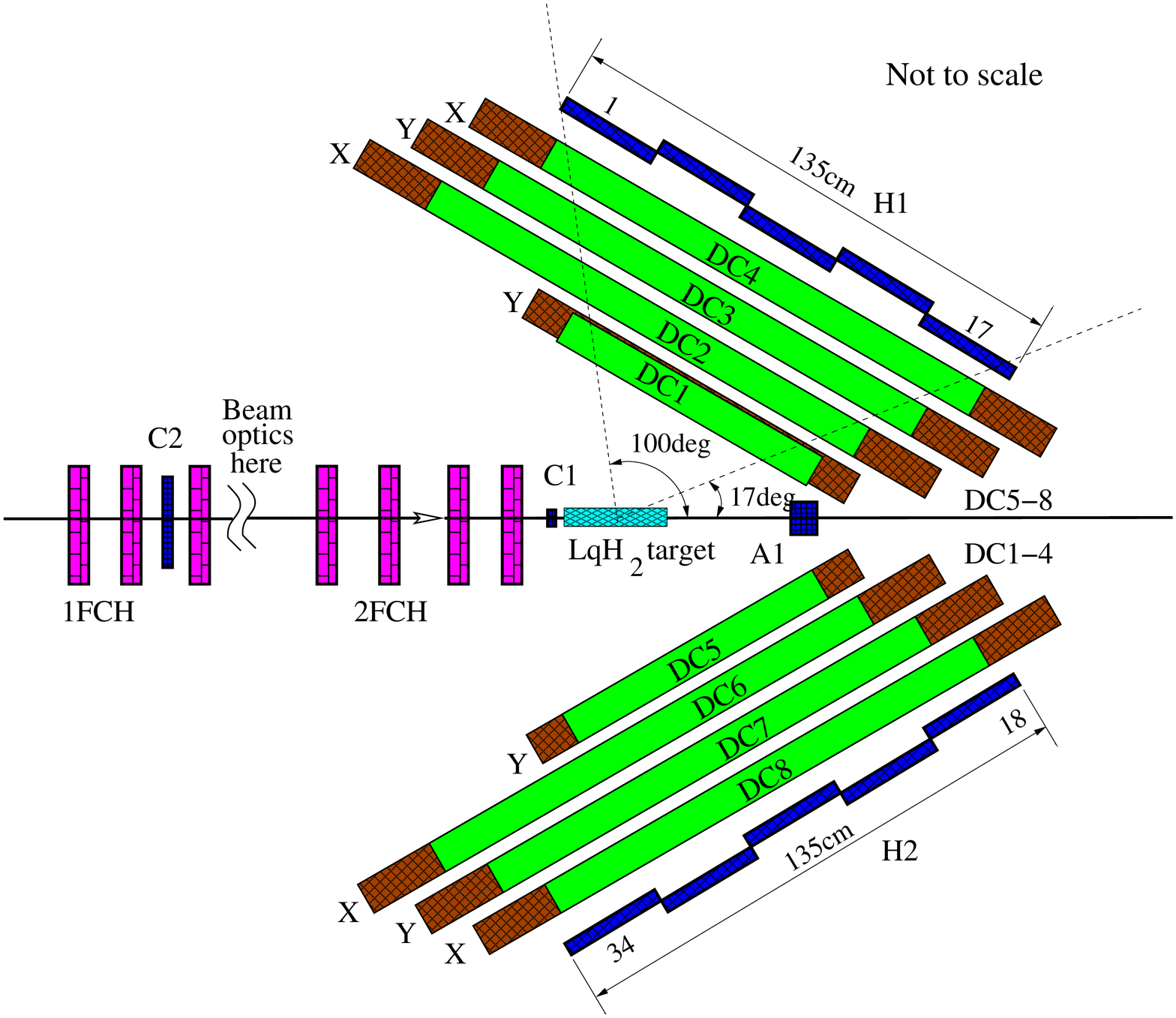}
    \vspace*{-6mm} %% the vertical position may need tweaking
    \caption{\footnotesize Experimental layout for $\pi^-p$ elastic scattering.}
    \label{alekseev_fig1}
    \vspace*{-6mm} %% the vertical position may need tweaking
\end{figure}

   The layout dedicated to the elastic scattering measurement is shown in fig.~\ref{alekseev_fig1}. The main parts are:
proportional chambers (\textbf{1FCH} and \textbf{2FCH}), drift chambers (\textbf{DC1-8}), liquid hydrogen target
(\textbf{LqH}$_2$), scintillation counters (\textbf{C1}, \textbf{C2} and \textbf{A1}) and two scintillation hodoscopes
(\textbf{H1} and \textbf{H2}). 

%%%%%%%% Fig. 1 %%%%%%%%

Proportional chambers are placed in the 1$^\mathrm{st}$ and 2$^\mathrm{nd}$ focuses of the beam. 
The chambers are two-coordinate, have square sensitive region of $200\times200$~mm$^2$, 
1~mm signal wires pitch, 40~um aluminum foil cathode electrodes and 6~mm between the foils.
We use "magic" gas mixture (argon-isobutane-freon) to feed the proportional chambers.
Beam tests showed efficiency better than 99\%. 

\begin{wrapfigure}[15]{R}{90mm}
    \centering
    \vspace*{-8mm} %% the vertical position may need tweaking
    \includegraphics[width=90mm]{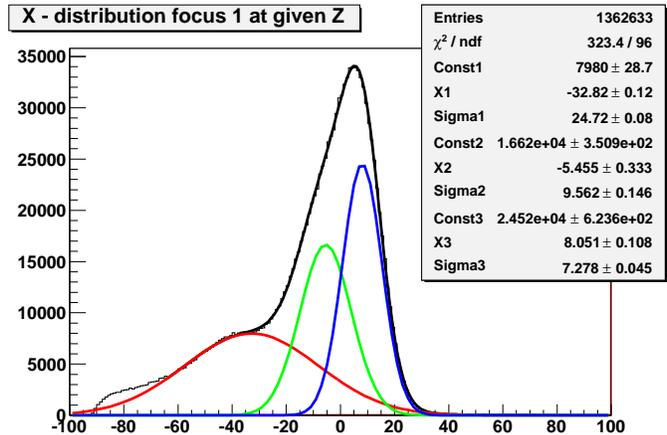}
    \vspace*{-7mm} %% the vertical position may need tweaking
    \caption{\footnotesize Horizontal distribution in the first focus of internal the accelerator 
    beam protons scattered over internal beryllium target.}
    \label{alekseev_fig2}
\end{wrapfigure}

Main task of the chambers in the 1$^\mathrm{st}$ focus is to measure the momentum of each pion
going to the target. Strong dipole magnets between the 
internal target and the 1$^\mathrm{st}$ focus provide horizontal distribution of the
particles with different momentum with dispersion 57~mm/\%. A distribution 
over horizontal coordinate in the 1$^\mathrm{st}$ focus of the events
of scattering of the internal beam protons with momentum 1.0~GeV/c over a beryllium 
internal target is shown in fig.~\ref{alekseev_fig2}. The peaks observed in the
picture correspond to (right to left) the elastic scattering, the first excitation of beryllium nucleus
and the second and the third excitations seen as one peak.

%%%%%%% Fig. 2 %%%%%%%%%%%%

The liquid hydrogen target has a mylar cylinder container with diameter 40~mm and the length about 250~mm placed 
in high vacuum inside beryllium outer shell 1~mm thick. It is connected by two pipes 
to the liquefier system. One is used for liquid hydrogen inflow and through the other the evaporated gas gets back
to liquefier. This design provides minimum of matter for the particles. The refrigeration is provided by liquid
helium, which flow is controlled by the feedback supporting constant pressure of the hydrogen in the 
closed volume. This pressure corresponds to proper ratio between liquid and gas fractions of the hydrogen
and thus ensuring that the liquid occupies whole target working volume and that the hydrogen is not frozen.
Pressures and temperatures in the target system are monitored and logged.

\begin{wrapfigure}[11]{R}{90mm}
    \centering
    \vspace*{-8mm} %% the vertical position may need tweaking
    \includegraphics[width=90mm]{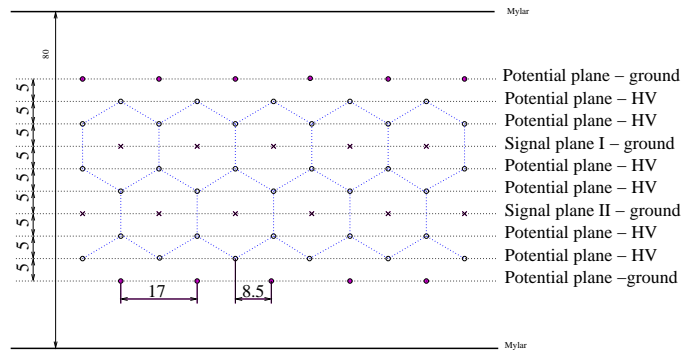}
    \vspace*{-7mm} %% the vertical position may need tweaking
    \caption{\footnotesize Drift chamber cross section. View along the wires.}
    \label{alekseev_fig3}
\end{wrapfigure}

There are 8 one coordinate drift chambers in the elastic setup. 6 chambers have sensitive region 
$1200\times 800$~mm$^2$ and for 2 chambers closest to the target it is $600\times 400$~mm$^2$.
The chambers have double sensitive layers hexagonal structure 
shown in fig.~\ref{alekseev_fig3}. Comparing to the conventional drift tubes this structure has much more 
complex fields, but provides significantly less amount of matter on the particle path. Potential wires form nearly 
regular hexagon with a side of 10~mm. Drift chambers are fed with 70\% Ar and 30\% CO$_2$ gas mixture. Beam
tests showed better than 99\% single layer efficiency and about 0.2~mm resolution.

%%%%%%% Fig. 3 %%%%%%%%%%%%

A unique distributed DAQ system based on the commercial 480~Mbit/s USB~2.0 interface was designed for the experiment
\cite{EPECURDAQ}. It consists of 100-channel boards for proportional chambers and 24-channel boards for drift chambers, 
placed on the chambers frames. Each board is connected by two cables (USB 2.0 and power) 
to the communication box, placed near the chamber. Then the data is transferred to the main 
DAQ computer by the standard TCP/IP connection. 
Trigger logic is capable of processing of several trigger conditions firing different sets of detectors. 

\begin{wrapfigure}[15]{R}{60mm}
    \centering
    \vspace*{-8mm} %% the vertical position may need tweaking
    \includegraphics[width=60mm]{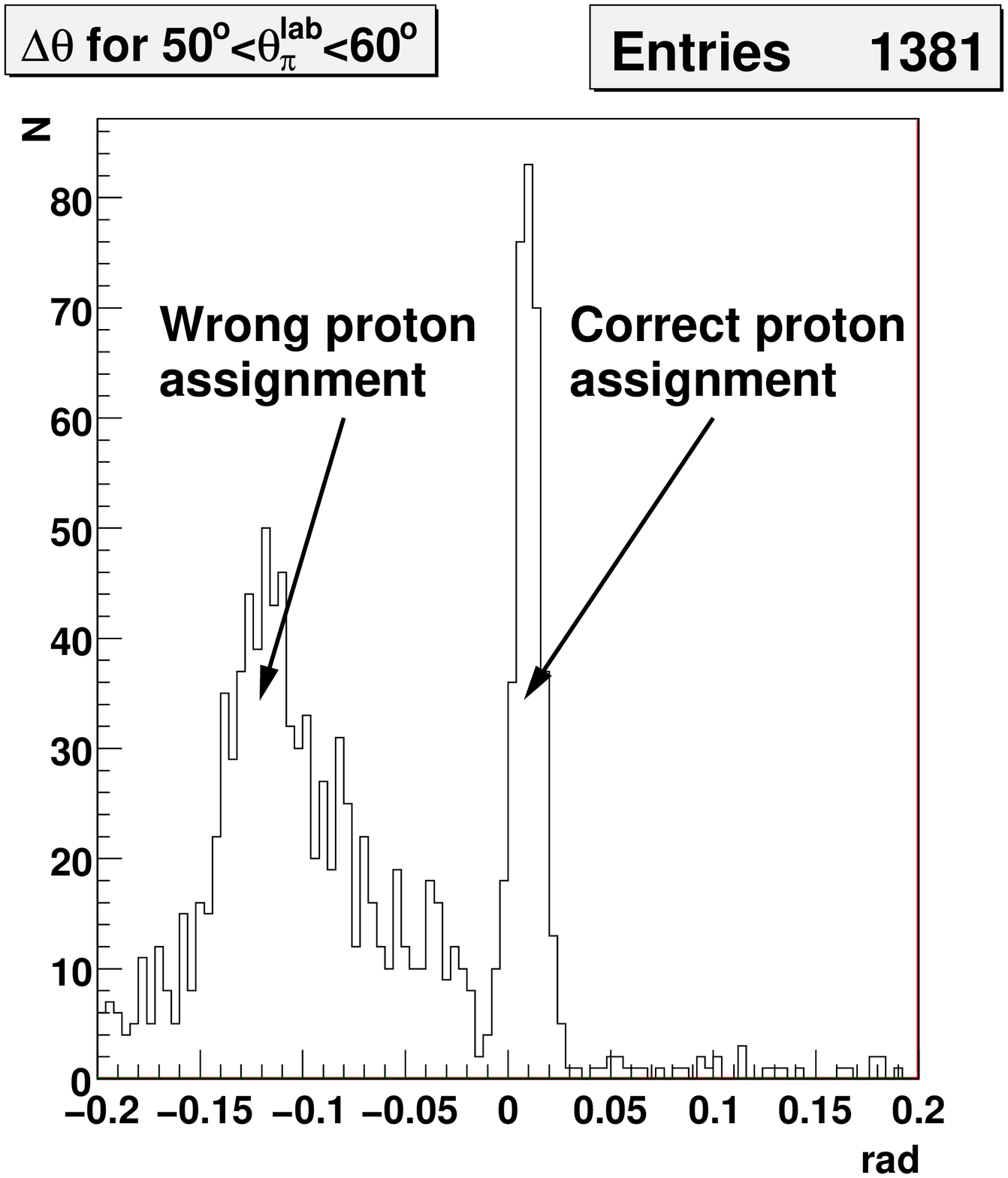}
    \vspace*{-10mm} %% the vertical position may need tweaking
    \caption{\footnotesize $\Delta\theta^\mathrm{CM}$ distribution for $\theta_\mathrm{lab}^\pi$ range $50^o - 60^o$.}
    \label{alekseev_fig4}
\end{wrapfigure}

During the engineering run last December and the first physics run in April this year the main trigger was set as: 
$$
T = C_1 \cdot C_2 \cdot M_{1FCH} \cdot M_{2FCH} \cdot \overline{A_1}
$$
where $C_1$, $C_2$ and $A_1$ - signals from corresponding scintillator counters and
$M_{1FCH}$ and $M_{2FCH}$ - majority logic of the proportional chamber planes in the 1$^\mathrm{st}$ and
the 2$^\mathrm{nd}$ focuses. Other trigger conditions were used to provide beam position and luminosity
monitoring. To ensure stable beam momentum an NMR monitoring of the magnetic field of the last dipole
was used. We collected over $5\cdot 10^8$ events in the April run, of which we expect about 5\% to be elastic
events, in the invariant mass range 1640--1745~MeV. Processing of the April run data had started. An example
of the elastic events selection for $50^o < \theta_\mathrm{lab}^\pi < 60^o$ is presented in fig.~\ref{alekseev_fig4}.
A distribution of the polar angles difference between pion and proton in the center of mass system is shown. 
A narrow peak corresponds to the correct assumption which of the secondary particles is pion and which is proton. 
A wide peak correspond to the wrong assumption. This comes from the fact that in the elastic setup we can't
distinguish between pion and proton and can only try some assignment and correct events for which it turned
out to be wrong. 

%%%%%%% Fig. 4 %%%%%%%%%%%%

As a conclusion:
\vspace*{-2mm}
\begin{itemize}
\itemsep=0mm
\parsep=0mm
\parskip=0mm
\item Construction of the first stage (elastic) of EPECUR has finished and the setup was successfully commissioned.
\item Data taking has started this year and the first $5\cdot 10^8$ events were collected.
\item Data processing is under way.
\item We are going to take more runs.
\item We are going to proceed to the construction of the second stage (K$\Lambda$-production) simultaneously with data
taking.
\end{itemize}
\vspace*{-2mm}

The work is supported by Russian Fund for Basic Research
(grants 05-02-17005-a and 09-02-00998-a) and Russian State Corporation
on the Atomic Energy 'Rosatom'.

\end{document}